\documentclass[adp,letterpaper,fleqn]{w-art}

\usepackage{graphicx}
\usepackage{slashed}
\usepackage{cite}
\usepackage{amssymb}
\usepackage{hyperref}

\newcommand{\nco}{\newcommand}
\nco{\beq}{\begin{equation}} \nco{\eeq}{\end{equation}}
\nco{\beqa}{\begin{eqnarray}} \nco{\eeqa}{\end{eqnarray}}
\nco{\lra}{\leftrightarrow}
\nco{\sss}{\scriptscriptstyle} \nco{\dphi}{\varphi}
\nco{\etal}{\textit{et al.}}
\nco{\VEV}[1]{\langle #1 \rangle}
\nco{\wt}{\widetilde}
\nco{\pref}[1]{(\ref{#1})}
\nco{\pa}{\phantom{2}}
\nco{\comment}[1]{}

\begin{document}
\DOIsuffix{} \Volume{} \Month{} \Year{} \pagespan{1}{}
\Receiveddate{}\Reviseddate{}\Accepteddate{}\Dateposted{}


\title[DM and 511 keV $\gamma$]{Abelian dark matter models for
511 keV $\gamma$ rays and direct detection}
\date{\today}

\author[J.M. Cline]{James M.~Cline\inst{1}
\footnote{email:~\textsf{jcline@physics.mcgill.ca}}}
\address[\inst{1}]{Department of Physics, McGill University, 
Montr\'eal, QC, H3A 2T8, Canada}

\author[A.R. Frey]{Andrew R.~Frey\inst{2}
\footnote{Corresponding author\ email:~\textsf{a.frey@uwinnipeg.ca},
phone: +1-204-786-9215, fax: +1-204-774-4134}}
\address[\inst{2}]{Dept.~of Physics and Winnipeg Institute for Theoretical 
Physics,
University of Winnipeg, Winnipeg, MB, R3B 2E9, Canada}

\begin{abstract}

We construct a simple  U(1) hidden sector model of
metastable dark matter that could explain excess 511 keV gamma rays
from the galactic center as observed by INTEGRAL, through inelastic
scattering of dark matter followed by its decay.  Although the model is highly
constrained, it naturally accommodates dark matter with mass and cross
section in the range suggested by the CoGeNT and CRESST experiments. 
The dark gauge boson that mediates the interactions with standard
model matter has a mass of several hundred MeV, and might be
discovered by heavy photon detection experiments, including APEX,
MAMI and HPS.

\end{abstract}

\maketitle

\tableofcontents

\section{Introduction}

While direct detection of dark matter (DM) is presently an  urgent
endeavor, it is possible that indirect signals of DM have already been
seen.  Most recently the detections of  excess high-energy 
galactic  electrons and
positrons by the PAMELA \cite{pamela} and Fermi/LAT \cite{fermi} experiments
 have stimulated such interest (although this signal is now generally
believed to be due to pulsars), 
but there is a much longer-standing observation originally from
\cite{jhh1972,lms1978} 
of excess 511 keV gamma rays from the galactic center (studied in the
greatest detail by INTEGRAL/SPI \cite{spi1,spi2,spi3,spi4}),
which might be explained by positrons from DM decays.  This
possibility is especially interesting because there is no widely
accepted astrophysical explanation for the observation 
\cite{astro1,astro2,astro3,astro4,astro5,astro6,astro7,astro8,astro9,astro10,
astro11}.  

Proposed astrophysical explanations for the excess 511 keV gamma rays 
include pulsars, X-ray binaries, violent events near
the galactic black hole (Sgr A*), and 
radionuclides from supernovae and massive stars; \cite{Prantzos:2010wi}
gives a thorough review and critical appraisal of these proposals, including 
DM proposals, as well as a discussion of the effects of positron 
propagation.  An important morphological feature of the 511 keV emission,
which is difficult to reproduce with steady-state astrophysical models, is
that the galactic bulge is several times brighter than the disk.
As an example, consider positron production by radionuclides, which has
been claimed as an explanation of the entire 511 keV signal by
\cite{astro9,astro10} based on a particular model of radionuclide production
and positron production with very specific assumptions.  
However, \cite{Martin:2012hv} recently argued
that, for updated models of radionuclide production, reasonable assumptions
about positron production cannot reproduce the observed bulge-to-disk ratio.
Therefore, while radionuclides can explain the observed disk emission, 
an additional positron source seems necessary to explain the bulge emission.\footnote{
To quote ref.\ \cite{Prantzos:2010wi}:
``However, in view
of the different conditions---density, magnetic field and
unknown level of small-scale turbulence---in the Bulge,
such a calculation [that of ref.\ \cite{astro9}] appears rather arbitrary (although not
necessarily wrong). Their model can be considered as a
quantitative illustration of a possible scenario, the plausibility 
of which remains to be shown.''  The criticisms of ref.\ 
\cite{Martin:2012hv} are more pointed: ``[The authors of ref.\ \cite{astro9}] 
claim that differential propagation of nucleosynthesis positrons can explain all 
properties of the INTEGRAL/SPI
observations. We reach a different conclusion, at least for the
morphology of the annihilation emission. We identified several [four]
assumptions in the work of [ref.\ \cite{astro9}]  
that may explain the discrepancy and review them in the following\dots"
Detailed criticisms of the claims in  \cite{astro10} that dark matter annihilation
cannot explain the observed signal are also given in section V.D of 
\cite{CFC}.}
We are therefore motivated to explore alternative proposals for the source 
of the bulge positrons, especially 
given the strong concentration of the signal in the inner bulge
\cite{Vincent:2012an}.

Although many DM models have been constructed to produce excess  positrons
through  inelastic scattering followed by decay 
\cite{Finkbeiner:2007kk,Pospelov:2007xh,ArkaniHamed:2008qn,Finkbeiner:2009mi,Chen:2009dm,
Chen:2009av,CFC,Morris:2011dj}, by  direct decays  of metastable DM
\cite{Picciotto:2004rp,Pospelov:2007xh,CFC}, or by direct annihilation of MeV-scale
DM \cite{Boehm:2003bt,Hooper:2003sh,Huh:2007zw}, it is important to note
whether
such models can be tested (and potentially excluded)
through complementary experiments.  Connecting to direct detection of the DM
is clearly interesting for any such scenario.  But in hidden sector
models, the new particles that mediate the DM interactions can also be
discoverable in beam dump experiments.  Moreover if the mediator is a scalar that
mixes with the standard model (SM) Higgs boson, then new signatures can arise in rare
decays involving the scalar.   Of the three classes of models mentioned above,
decaying DM is ruled out as a source of galactic 511 keV gamma rays because it
predicts a much broader spatial distribution than is observed
\cite{Boehm:2010qt,Abidin:2010ea}.  Inelastic scattering to excited dark matter
(XDM) on the other hand gives a very good fit to the observed shape
\cite{Vincent:2012an}.  In addition, the mass scale of XDM is not restricted,
since it only requires that the mass splitting between nearby states, not
the masses themselves, be at the MeV scale.  This affords stronger 
possibilities for such DM to be discovered in existing or imminent direct detection 
experiments.

In the present work, we propose a new class of XDM  models, guided by the
desire for simplicity and for complementary signals in a range of experiments,
which may exclude or constrain
these models.  Many of the recent XDM models were based upon
exchange of hidden sector nonabelian gauge bosons
\cite{ArkaniHamed:2008qn,Chen:2009dm,Chen:2009ab,CFC}.  In the present
work we aim to build the minimal XDM model based upon an Abelian gauge
symmetry.  It will be seen that these models are not as economical as
the original XDM  model proposed in  \cite{Finkbeiner:2007kk}, which
relied upon scalar boson exchange. The main reason for additional
complications in the case of gauge boson exchange is that DM
scatterings must be inelastic at {\it both} interaction vertices,
whereas with scalar exchange, it is possible to have elastic
scattering at one vertex.  The result is that gauge models necessarily
produce two $e^+ e^-$ pairs, which is energetically more difficult
than making just a single pair, and suppresses the rate of positron
production since one relies upon the tail of the DM velocity distribution to
have enough kinetic energy to produce the pairs.  While it is possible to  achieve the observed rate in the
galactic center with scalar models having only two dark matter states \cite{Morris:2011dj}, it is much
more 
difficult to do so in gauge boson models \cite{Chen:2009av}.  To
overcome this difficulty, it was proposed to have three states of DM,
one of which is metastable and having a smaller energy gap to its
neighboring state than $2 m_e$ \cite{Chen:2009dm}.  The smaller gap
allows for a larger rate of excitation, or de-excitation in case the
metastable state is the heaviest of the three \cite{CFC}.  Because of
this need for three DM states, our present model requires three Weyl
DM particles, and a kinetically mixed U(1) gauge boson. It also
requires two new dark scalars, one real and one complex.  Despite the
many free parameters, we will show that the model is highly
constrained and allows for a number of complementary experimental
tests.  In contrast, the much simpler model of
\cite{Finkbeiner:2007kk} does not seem to offer any additional tests
that would facilitate its exclusion (or provide hints toward
its confirmation).

Our analysis is similar to that in ref.\ \cite{CFC}, except that the
latter considered nonabelian DM which required rather
elaborate Higgs sectors and hierarchies of gauge kinetic mixing  in
order to get the desired mass splittings and lifetime of the
metastable DM state.  The model presented here is 
simpler.  We also employ improvements in direct detection limits from
XENON100  and updated analyses of the CoGeNT  signals
relative to what existed at the time of ref.\ \cite{CFC}.

\section{Defining the model}

The minimal models for XDM consist of three Weyl
fermions as the DM states.\footnote{We prefer fermions to scalars 
for the DM because
the small masses and mass splittings of fermions are stable under
renormalization.}   Since the top two are nearly degenerate, it is
natural to consider them as being originally the components of a Dirac
fermion, split by some small symmetry breaking due to a hidden
sector complex Higgs $\phi$; after symmetry breaking, the DM is a triplet
of Majorana fermions $\chi_{1,2,3}$.  
As in \cite{CFC}, the mass splitting $\delta M_{12}$ between the lowest
and two highest states must be accidentally of order a few MeV, while the smaller splitting 
$\delta M_{23}$ between
the two highest states appears naturally due to the underlying symmetry.
We will also require an additional real singlet $\Phi$ in order to get the
correct relic density for the DM ground state $\chi_1$.  

We denote the three Weyl states in the Lagrangian basis by $\psi_1$, 
$\psi_2$, $s$, with charges $+1$, $-1$, $0$
respectively under the dark gauged U(1).  Let the scalars $\phi$ and $\Phi$ have 
charge $-1$ and 0 respectively.  Omitting for now the Higgs potential terms,
this gives rise to the Lagrangian
\beq
\mathcal{L} = 
M_\chi \psi_1 \psi_2 + \frac 12 (m+y_\Phi\Phi) s s 
+ (y_1\phi \psi_1  s 
+ y_2\phi^* \psi_2  s
+ \mathrm{h.c.}) +g\,Z'_\mu(\psi_1^\dagger \sigma^\mu \psi_1 - 
\psi_2^\dagger \sigma^\mu \psi_2)
\label{xdm_mass}
\eeq
where $Z'$ is the new U(1) gauge boson, and we use
the shorthand $\psi_1 \psi_2 = \psi_1^T\sigma_2\psi_2$ for the
Lorentz invariant spin contractions.  The $Z'$ is assumed to  kinetically mix
with the photon (properly hypercharge), and the two scalars will 
acquire some small mixing with the SM Higgs $H$, by virtue of Higgs portal terms
$|H|^2|\phi|^2$ and $|H|^2 \Phi^2$ and nonvanishing VEVs for $\phi$ and $\Phi$.
Furthermore, $\phi$ and $\Phi$ can mix with each other after symmetry breaking
through $|\phi|^2\Phi^2$; for simplicity, we will assume that this mixing angle is 
negligible in the following.
For notational convenience, we henceforth absorb
$y_\Phi\langle\Phi\rangle$ into the definition of $m$.

Suppose that the masses are such that $M_\chi - m \gtrsim 2 m_e$\footnote{
Only rarely do the positrons pair-annihilate in flight (they must first
form positronium), so the $e^+$ injection energy,
which is of order $M_\chi - m \gtrsim 2 m_e$, must be $\lesssim 10$ MeV;
see, for example, \cite{Prantzos:2010wi}.} 
(and provisionally $m,M_\chi \sim 10$ GeV to
eventually  make contact with direct detection experiments),
so that before $\phi$ gets a VEV, we have a Dirac state with slightly 
higher mass
than that of $s$.  When $\phi$ gets its VEV, this splits into two 
nearby Majorana states with masses
\beq\label{split}
      M_+ =   M_\chi +  \frac{(\mu_1 + \mu_2)^2}{2(M_\chi - m)},
	\quad M_- = M_\chi + \frac{(\mu_1 - \mu_2)^2}{2(M_\chi + m)}
\eeq
where $\mu_i \equiv y_i \langle\phi\rangle$.  Presently, we will argue
for a symmetry that imposes $\mu_1+\mu_2=0$.  
Then we can use the seesaw mechanism to explain the smaller of the
two mass splittings so long as $\mu_i \ll (M_\chi + m)$.  For example if
$\mu_i \sim$ 10 MeV, then we get $\delta M_{23}\sim 10$ keV
which is the desired scale indicated by our earlier study of a similar model
in ref.\ \cite{CFC}.  

We need to gauge the U(1) symmetry to get rid of the Goldstone boson
that would arise from $\phi$ getting its VEV; otherwise the massless
boson could mediate problematic long-range forces coupling to the 
dark matter.  This is the origin of the dark gauge boson $Z'$.
Notice that with our choice of charges the
model is anomaly free. In the limit of vanishing gauge kinetic
mixing, $\epsilon\to 0$, this is obvious because $Z'$ couples only to a Dirac fermion
$\psi$.
Nonvanishing $\epsilon$ can be induced for example by integrating out a heavy
scalar that carries both the dark U(1) and weak hypercharge, which cannot
induce any anomaly.  Ignoring the $Z$ boson, the gauge kinetic mixing term 
is
\beq
	-{\epsilon\over 2} F^{\mu\nu}Z'_{\mu\nu}
\eeq
where $F_{\mu\nu}$ is the electromagnetic field strength and $Z'_{\mu\nu}$
is that of the dark U(1).  After diagonalization of the kinetic term,
$Z'$ acquires a coupling to the electromagnetic current with the extra
$\epsilon$ suppression factor.  This allows $Z'$ to mediate the decay
$\chi_2\to \chi_1 e^+ e^-$.

\begin{figure}[t]
\sidecaption
\includegraphics[width=48mm]{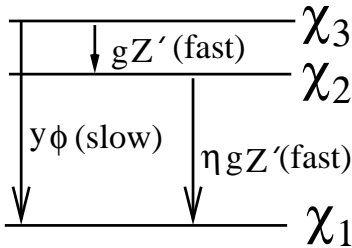}
\caption{Schematic depiction of DM mass spectrum and the transitions between the
states.  The $\chi_3\to\chi_1 e^+ e^-$ transition is a very slow decay mediated
by $\phi$, while
$\chi_3\chi_3\to\chi_2\chi_2$ is a fast
annihilation process and $\chi_2\to\chi_1 e^+ e^-$ is a fast decay
mediated by $Z'$.\bigskip\bigskip}
\label{spectrum}
\end{figure}

One of the challenges for this model is that the rates of
decay of the mass eigenstates $\chi_2\to \chi_1 e^+ e^-$ and 
$\chi_3\to \chi_1 e^+ e^-$ must be quite different from each other
in order for $\chi_3$ to be cosmologically long-lived while $\chi_2$
decays relatively promptly.  
This difference can be naturally realized
by invoking a discrete symmetry $\psi_1\leftrightarrow \psi_2$,
$\phi\leftrightarrow -\phi^*$ that insures that $y\equiv y_1 = -y_2$ hence
$\mu_1 = -\mu_2$.  In this case we obtain 
distinctly different couplings between the two highest mass 
eigenstates and the lowest one.  The relation between the flavor states and the
mass states is approximately
\beq 
\psi_1 \simeq \frac{1}{\sqrt{2}}(\chi_2 + \chi_3 + \eta\chi_1)\ ,\ \ 
\psi_2 \simeq \frac{1}{\sqrt{2}}(\chi_2 - \chi_3 - \eta\chi_1)\ ,\ \
s\simeq \chi_1 - \eta \chi_3\ ,
\label{mixing}
\eeq
where $\eta \equiv \sqrt{2}y\langle\phi\rangle/(M_\chi+m)$ is the 
small mixing parameter between the 
$\psi$ and $s$ sectors.  The mass splitting between the top two states
is $\delta M_{23} = \eta^2 (M_\chi+m)$.
With the above choices, we obtain different kinds of couplings between
$\chi_1$-$\chi_2$ and $\chi_1$-$\chi_3$,
\beq
\mathcal{L}_{\mathrm{int}}=
y\phi\left[ \chi_3 \chi_1 + \eta(\chi_1^2-\chi_3^2)\right] +
\frac 12 y_\Phi\Phi
\left[\chi_1^2-\eta(\chi_1\chi_3+\chi_3\chi_1)\right]+   
\chi_2^\dagger\, g\,Z'_\mu\sigma^\mu(\chi_3 +\eta\chi_1)
+{\rm h.c.}
\label{couplings}
\eeq
where now $\phi$ stands for the canonically normalized real part of
the complex scalar.  This allows us to create a natural hierarchy between the $\chi_2$ and $\chi_3$
lifetimes.  The decay $\chi_3\to\chi_1 e^+ e^-$ is mediated only by
$\phi$, and that of $\chi_2\to\chi_1 e^+ e^-$ proceeds by the $Z'$.
In terms of 4-component Majorana spinors $\chi_{1,2,3}$, the Yukawa 
and gauge couplings become
\beq \mathcal{L}_{\mathrm{int}}=
   y\phi\left[ \bar\chi_3 \chi_1 +\eta(\bar\chi_1\chi_1 -
   \bar\chi_3\chi_3) \right]+ y_\Phi\Phi\left(\bar\chi_1\chi_1 -\eta\bar\chi_1\chi_3
   \right)+g \bar\chi_2 \slashed{Z}' (\chi_3+
\eta\chi_1)\ .
\eeq
The mass spectrum and the different kinds of transitions between the states
are depicted qualitatively in fig.\ \ref{spectrum}.  The model is necessarily
{\it exothermic}: the heaviest state $\chi_3$ is the one that can be made metastable
on time scales greater than the age of the universe.    Thus the process
$\chi_3\chi_3\to\chi_2\chi_2$, that leads to fast $\chi_2\to\chi_1 e^+e^-$ decays
at the galactic center, is not impeded by any energy barrier.

\section{The 511 keV signal and relic density}

In this section, we will discuss requirements on the couplings of our
model to account for both the DM relic abundance and the 511 keV signal from 
the galactic center.  These are the two major constraints on any
XDM model which must be satisfied.\footnote{Another possibility, which we
do not consider here, is to get the 511 keV signal from a subdominant
component of dark matter whose excitation cross section is correspondingly
larger.}\ \   We consider them in turn.

\subsection{Scattering rate for $e^+$ production}
\begin{figure}[t]
\sidecaption
\includegraphics[width=68mm]{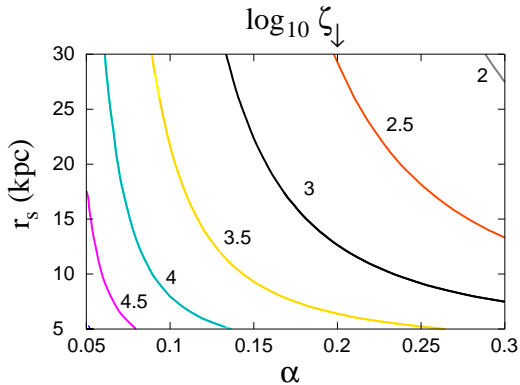}
\caption{Contours of the log of
$\zeta_{\downarrow}$, defined in (\ref{zetap_down}),
as a function of Einasto halo profile parameters $\alpha$ and $r_s$.
The threshold velocity $v_t = 1.4\times 10^{-3}$ is assumed.
\bigskip\bigskip\bigskip\bigskip\bigskip}
\label{zetap}
\end{figure}

The cross section for $\chi_3\chi_3\to\chi_2\chi_2$
is given in eq.\ (36) of 
ref.\ \cite{CFC}.  For the parameter values of interest here,
it simplifies to
\beqa
\sigma_{\downarrow}v_{\mathrm{rel}} &=&
\frac{g^4 M_\chi^2}{4\pi m_{Z'}^4}\sqrt{v^2+v_t^2}\ ,
\eeqa
where the $v_t$ denotes the threshold velocity for the {\it inverse}
reaction $\chi_2\chi_2\to\chi_3\chi_3$ in the center-of-mass
frame: $v_t = \sqrt{2\delta M_{23}/M_\chi}$.  The rate of
scatterings in the galactic center is given by eq.\ (28) of the same
reference: $R_{e^+} = 2\times \frac12 (Y_{3}/Y_{\mathrm{tot}})^2 \int d^{\,3}x
\langle\sigma_{\downarrow}v_{\mathrm{rel}}\rangle (\rho/M_\chi)^2$, where
$Y_{3}/Y_{\mathrm{tot}}$ is the relative abundance of $\chi_{3}$, $\rho$
is the DM mass density, and the integral is taken over radii
corresponding to the INTEGRAL observations, $r\lesssim r_c\equiv 1.5$ kpc
(the first factor of 2 accounts for the fact that each scattering produces 
two  $e^+$-$e^-$ pairs).
One must perform the phase space average over the DM
velocity distribution to compute $\langle\sigma_{\downarrow}v_{\mathrm{rel}}\rangle$, 
which depends upon $r$ because the mean and
escape velocities of the DM are $r$-dependent.   We have carried out
these integrations numerically for a range of different DM density
profiles, assumed to be Einasto profiles $\rho = \rho_\odot\exp(
-(2/\alpha)\left( (r/r_s)^\alpha - (r_\odot/r_s)^\alpha \right)$
with $\rho_\odot = 0.3$ GeV/cm$^3$.
In this way we can express the rate of positron production in the form
\beq
R_{e^+} = 4\pi\,\zeta_{\downarrow} v_t\frac{\rho_\odot^2}{M_\chi^2} 
\left(\frac{Y_{3}}{Y_{\mathrm{tot}}}\right)^2 \overline{\sigma v}
\times (1\textnormal{ kpc})^3\label{xdmrate}
\eeq
where $\overline{\sigma v} = {g^4 M_\chi^2/ 4\pi m_{Z'}^4}$ and
\beqa
\label{zetap_down}
\zeta_\downarrow &=& \textnormal{kpc}^{-3}\int_0^{r_c} dr\, r^2 (\rho/\rho_\odot)^2
\left\langle\sqrt{v^2/v_t^2+1}\right\rangle\ .
\eeqa
We have used the same velocity distribution function as described in
\cite{CFC} for the phase-space average, and assumed that $v_t =2 \eta=
\sqrt{2\times 10^{-6}}$, corresponding to $\delta M_{23}=10$ keV,
$M_\chi = 10$ GeV.  The numerical results for
$\zeta_{\downarrow}$ are shown in 
fig.\ \ref{zetap}.  Equating (\ref{xdmrate}) to the observed rate
of $1.1\times 10^{43}$ s$^{-1}$ gives the constraint
\beq
\frac{m_{Z'}}{g} = 6.5 \textnormal{ GeV}\,
\left(\frac{\zeta_{\downarrow}}{3000}\right)^{1/4}
\left(\frac{\eta}{10^{-3}}\right)^{1/4}
\left(\frac{\bar Y}{0.33}\right)^{1/2}\ ,
\label{xdmscatt}
\eeq
where we defined $\bar Y = Y_3/Y_{\rm tot}$.
We note that for DM halo profiles close to those found in N-body
simulations for Milky Way-like galaxies, $\zeta_{\downarrow}\sim 3000$,
leading to the restriction that $m_{Z'} \lesssim 20$ GeV to keep the coupling
$g$ perturbative ($g\lesssim \sqrt{4\pi}$).  We find that the contours of 
 $\zeta_\downarrow$ shift very little for values of $v_t = 2\eta$ larger than the
assumed $1.4\times 10^{-3}$,
but they start to shift noticeably to the right for smaller values.

The total positron production rate within the bulge is, of course, not the 
only observable characterizing 
the 511 keV gamma ray signal.  As has been stressed in
\cite{Boehm:2010qt,Abidin:2010ea,Vincent:2012an}, the gamma ray intensity is strongly
peaked near the galactic center, with a best fit similar to $\rho(r)^2$ for 
an Einasto profile with $\alpha=0.17$, $r_s=26$ kpc \cite{Vincent:2012an}.  
While it is possible
that positrons from outside the center of the galaxy propagate toward the center to
annihilate \cite{astro9,astro10}, it seems simpler to assume that the $e^+$ sources 
are concentrated in the galactic center and that $e^+$ do not propagate long distances
before annihilating.  In fact ref.\ \cite{Martin:2012hv} has recently argued that,
while positrons may escape the galactic bulge or disk in some propagation models,
those which annihilate tend to do so near their sources, at least for
positrons that are injected with only 
several hundred keV of kinetic energy.  It therefore
seems reasonable to assume that the positrons do not travel very far before
annihilating.

\subsection{Relic density}

The requirement of getting the right DM relic density through thermal
decoupling significantly restricts the parameters of the model.   We start
with a summary of the key considerations. As we will show in section
\ref{washout}, the $\chi_1$ and  $\chi_{2,3}$ states do not scatter
efficiently into each other after  the DM species freeze out chemically, and
scatterings of $\chi_{2,3}$ into each other freeze out before $\chi_3$ is
significantly depleted.   As a result, chemical freezeout determines both the
total DM relic density and the relative abundance $Y_3$ of the metastable DM
state.  Moreover we will show that at the time of  chemical freezeout,
$\chi_2$ and $\chi_3$ have the same abundance; in order for them to be the main
components of the DM, we require the abundance of $\chi_1$ to be subdominant.
However $\chi_1$ couples too weakly to $Z'$ and $\phi$ to efficiently 
annihilate, so we take advantage of the singlet Higgs $\Phi$ to dilute it
through $\chi_1\chi_1\to\Phi\Phi$.
On the other hand, $\chi_2$ and $\chi_3$ have
an unsuppressed coupling to $Z'$, so the gauge interactions dominate their
annihilation cross sections and determine their relic density.  A summary of
the values of the couplings and masses needed to satisfy these restrictions
(as well as others to be discussed below) is provided in 
table \ref{parameters}.  Typically the scalars $\phi,\Phi$ must be somewhat lighter than
the DM, while the $Z'$ may be lighter or more massive.

\begin{table}[t]
\tabsidecaption
\begin{tabular}{|c|c|c|}
\hline
\textbf{Parameter} & \textbf{Scaling} & \textbf{Fiducial Value}\\
\hline
$M_\chi$ & independent & 10{\rm\ GeV} \\
$\eta$ &.. & $10^{-3}$ \\
$\zeta_\downarrow$ &.. & 3000 \\
$\epsilon$ &.. & $\gtrsim 10^{-6}$\\
\hline
$\mu_\Phi\sim m_\Phi$ & constrained & $\gtrsim 5$ GeV\\
$y_\Phi$ &..  & $\lesssim 1/10$\\
$\Theta$ &..  & $\lesssim 6\times 10^{-4}$\\
\hline
$m_\phi\sim\langle\phi\rangle$ & 
$(\zeta_\downarrow \eta)^{1/4}\bar Y^{1/2}$ & 4.6 GeV\\
$y$ & $M_\chi \eta^{3/4}\zeta_\downarrow^{-1/4} \bar Y^{-1/2}$ & $3\times 10^{-3}$\\
$\theta$ & $(M_\chi/\bar Y)^{1/2}$& $\lesssim 10^{-5}$\\
\hline
$m_{Z'}$ ($\chi\chi\to Z'Z'$) & $(\zeta_\downarrow\eta \bar Y)^{1/4} M_\chi^{1/2}$&
610 MeV\\
$g$ & $M_\chi^{1/2} \bar Y^{-1/4}$& 0.094\\
\hline
$m_{Z'}$ ($\chi\chi\to f\bar f$) & $M_\chi$ & 20 GeV\\
$g$ & $M_\chi (\zeta_\downarrow \eta)^{-1/4} \bar Y^{-1/2}$ & 3.08\\
\hline
\end{tabular}
\caption{Model parameters and their fiducial values.
The second column shows how the fiducial value of the 
dependent parameters
scales with the values of independent ones.
$m_{Z'}$ and $g$ are given for two possible dominant
annihilation channels.}
\label{parameters}
\end{table}

We consider first the total relic abundance of $\chi_{2,3}$, which is 
determined by the dark gauge interactions.
The strongest annihilation channels available are $\chi_2\chi_2\to
Z' Z'$, $\chi_3\chi_3\to Z' Z'$ and $\chi_2\chi_3\to f\bar f$ where
$f$ is a SM fermion.  $\chi_2$ and $\chi_3$ can also annihilate into the scalars
$\phi,\Phi$, but the relevant couplings are suppressed, so we can ignore
these channels.
The average annihilation cross section for $\chi_2\chi_2$, $\chi_2\chi_3$,
$\chi_3\chi_3$, including the annihilations into
$Z'$ and co-annihilations into SM fermions, is given by
\beq
\langle \sigma_{\textrm{ann}} v\rangle = \frac{1}{32\pi\, M_\chi^2}\,
S(g,\epsilon,m_{Z'}/M_\chi)\ ,
\label{sigv}
\eeq
where
\beq
S(g,\epsilon,x) = \frac 12\, g^4\, f_1(x) +
(g\epsilon e)^2 \sum_i  N_{c,i}\, Q_i^2\, f_2(x,x_i)\ ,
\label{Xeq}
\eeq
with the sum over all kinematically allowed SM fermions having charge 
$Q_i$ and number of 
colors $N_{c,i}$.  The kinematic functions are given by 
$f_1 = (1-x^2)^{3/2} /\left(1-\frac12 x^2\right)^2\,\Theta(1-x)$
and $f_2(x,x_i) = \left(1+\frac12 x_i^2\right)(1-x_i^2)^{1/2}
/\left(1-\frac14 x^2\right)^2\,\Theta(1-x_i)$, where $x_i = m_i/M_\chi$.
This cross section controls the total abundance of $\chi_2$ and $\chi_3$.
To analyze this, we eliminate $g$ in favor of $m_{Z'}$ using
(\ref{xdmscatt}) and taking our canonical values for $\eta$, $\zeta_\downarrow$,
$\bar Y$ (see table 1).  For several values of $\epsilon$, we determine 
contours in the plane of $M_\chi$-$m_{Z'}$ that give the required
cross section, $\langle \sigma_{\rm ann} v\rangle
= (2\bar Y)^{-1}\times 3\times 10^{-26}$ cm$^3$/s.  The factor of $2\bar Y$ reduces
the relative abundance of $\chi_3$ to $\bar Y$, with the same relative 
abundance for $\chi_2$ at freezeout (since annihilations of $\chi_2$ have the same
cross section as annihilations of $\chi_3$); at generic points in the allowed
parameter space, there is no kinetic equilibrium after freeze-out, so 
the present-day value is the same as that at the freeze-out temperature,
$\bar Y_0=\bar Y(T_f)$.
(More quantitatively, the required cross section is
reduced slightly due to the presence of extra light degrees of freedom;
see, for example, the discussion of \cite{CFC}.)

\begin{figure}[t]
\sidecaption
\includegraphics[scale=0.5]{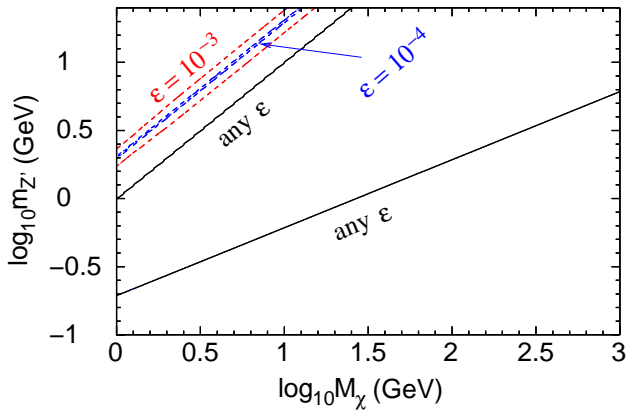}
\caption{Contours corresponding to the 
observed relic density for
$\chi_{2,3}$ in the $M_\chi$-$m_{Z'}$ plane, assuming the relation
(\ref{xdmscatt}) and $\zeta_{\downarrow} = 3000$.  Lowest two curves
(solid) are due to $\chi\chi\to Z'Z'$, while higher ones (dashed) are
due to $\chi\chi\to f\bar f$.
\bigskip\bigskip\bigskip\bigskip\bigskip}
\label{relicS}
\end{figure}

The resulting possible relations between $M_\chi$ and $m_{Z'}$ giving the  
observed relic density are  shown in fig.\
\ref{relicS}, for several values of the gauge kinetic mixing parameter $\epsilon$.
For any value of $\epsilon$, there are always two
solutions in which the $g^4$ contribution to $\sigma_{\rm ann}$
dominates.  The lower branch has
\beqa 
m_{Z'}&=& ( 610\ \textnormal{MeV})\times \left(\frac{M_\chi}{10\
\textnormal{GeV}}\right)^{1/2} \left(\frac{\zeta_\downarrow}{3000}\right)^{1/4}
\left(\frac{\eta}{10^{-3}}\right)^{1/4} \left(\frac{\bar Y}{0.33}\right)^{1/4}
\nonumber\\
g&=& 0.094\left(\frac{M_\chi}{10\ \textnormal{GeV}}\right)^{1/2} \left(
\frac{\bar Y}{0.33}\right)^{-1/4}\ .\label{relicvals}\eeqa
The upper branch is the unlikely situation in which $g$ would be too
large to give the right relic density, but $m_{Z'}\simeq M_\chi$ so that the
phase space nearly vanishes; we will not explore this case.  
There are also solutions where annihilation to SM fermions dominates at
$m_{Z'}>M_\chi$.  These contours depend somewhat on the value of $\epsilon$,
$\eta$, and $\zeta'_\downarrow$,
but they are centered around the condition for resonance in the $s$-channel,
\beqa m_{Z'}&\simeq& 2 M_\chi\nonumber \\
g&\simeq& 3.08 \left(\frac{M_\chi}{10\ \textnormal{GeV}}\right)
\left(\frac{\zeta_\downarrow}{3000}\right)^{-1/4}
\left(\frac{\eta}{10^{-3}}\right)^{-1/4}\left(\frac{\bar Y}{0.33}\right)^{-1/2}\ .
\eeqa
The gauge coupling tends to be nonperturbatively large on this branch,
unless $M_\chi$ is significantly smaller than 10 GeV.

In the case that $\chi\chi\to Z'Z'$ determines the $\chi_{2,3}$ abundance,
it is worth noting that $\chi_2$, $\chi_3$, and $Z'$ were 
in equilibrium with the SM until the time of freeze-out, so that the standard
estimate of their relic abundance applies.   To see this, consider
the Compton-like scattering $Z' f\to \gamma f$
($Z'\gamma\to f\bar f$ pair production has a similar cross section), 
where $f$ is a relativistic
SM fermion; at the temperatures when $\chi_{2,3}$ freeze out, these are $e,\mu,u,d,s$.
Ignoring all particle masses for simplicity and using $3T$ as an estimate for
typical energies and momenta, we estimate the cross section for this 
scattering as
\beq \langle\sigma_{Z'f\to\gamma f} v\rangle \approx 
\frac{\pi Q_i^4\alpha^2\epsilon^2}{18T^2}\ .\eeq
If we sum over all the available fermion species (including colors), 
the rate per $Z'$ to scatter into
$\gamma$ is larger than $3H$ at $T\approx 500$ MeV for $\epsilon\gtrsim 3\times 10^{-7}$.
As we will see in section \ref{lifetimes} below, $\epsilon$ must be larger than
this value for $\chi_2$ to decay quickly enough.  In fact, we have checked
in detail (including all particle masses and appropriate thermal averages) 
that this scattering remains in equilibrium down to temperatures of about
the electron mass for $\epsilon\gtrsim 10^{-6}$.  Therefore, the dark sector
remains in equilibrium with the SM through $\chi_{2,3}$ freezeout.

For the relic density of $\chi_1$, the  coannihilation
$\chi_1\chi_2\to  Z' Z'$ has an amplitude  suppressed by $\eta\sim
10^{-3}$, so there must exist other annihilation channels to prevent
$\chi_1$ from dominating the DM density and thereby diluting the
contribution of $\chi_3$ to the 511 keV signal. Similarly, as  will be
shown in sections \ref{relations} and 
\ref{constraints},
annihilations to $\phi$ are suppressed by the small  Yukawa
coupling $y$.  If the real singlet $\Phi$ is light however, we can
consider  $\chi_1\chi_1\to\Phi\Phi$ annihilations.   (There are also
annihilations $\chi_1\chi_1\to f\bar f$ to light SM fermions by 
$s$-channel boson exchange, but this is greatly suppressed by the
small  SM Yukawa couplings and $\Phi$-Higgs mixing angle.)
Annihilation to $2\Phi$ proceeds by  $s,t,u$ channels with a trilinear
$\Phi$ coupling $\mu_\Phi$  in the $s$ channel, and the cross section
is $p$-wave suppressed.  When the $s$-channel dominates, 
\beq 
   \sigma v_{\rm rel} \approx 
   \frac{\mu_\Phi^2 y_\Phi^2}{512\pi M_\chi^4} v^2 \frac{\sqrt{1-x^2}}{(1-x^2/4)^2}
\ ,\ \ x=m_\Phi/M_\chi\ .
\eeq
Assuming that freezeout occurs at $T\sim M_\chi/18$ (as is appropriate for
DM with mass $\sim 10$ GeV) and 
demanding that  $\langle \sigma_{\rm ann} v\rangle =  
3\times 10^{-26}\,Y_{\rm tot}/Y_1$
cm$^3$/s  to obtain $\chi_1$ relic abundance $Y_1<Y_{\mathrm{tot}}$ 
leads to the constraint 
\beq 
\mu_\Phi \simeq  \frac{0.15}{y_\Phi}\, \left(\frac{Y_{\rm tot}}{Y_1}\right)^{1/2}
\left(\frac{M_\chi}{10\textnormal{ GeV}}\right)^2\ \textnormal{GeV}\ ,
\label{lambda3}
\eeq 
where we have assumed 
that the kinematic
$x$-dependent factors are order unity.   We can achieve an acceptable
abundance, $Y_1/Y_{\rm tot}=0.006$ for example, with 
reasonable parameter choices like $y_\Phi = 0.1$,  $\mu_\Phi\simeq
m_\Phi\simeq 5$ GeV.  This relationship could also hold for annihilation 
$\chi_1\chi_1\to \phi\phi$ if $\mu_\Phi$ is the coupling for $\Phi|\phi|^2$.

We note that the value of $Y_1$ at freezeout is not the same as its
present-day value; $\chi_2\to \chi_1 e^+ e^-$ decays will enhance $Y_1$
nonthermally, as the rate of this decay is generically much slower than the
Hubble rate at chemical freezeout.   
Since only $\chi_3$ participates in $e^+$ production
and direct detection and $Y_3$ is fixed after chemical freezeout, we are
not concerned with the present-day value of $Y_1$.

\subsection{Relations between model parameters}\label{relations}

In addition to the relationships between model parameters required for the 
observed relic density and galactic positron production, there are 
several others.

The complex scalar vacuum expectation value and $Z'$ mass satisfy 
\beq \frac 12 m_{Z'}^2 \leq g^2 \langle\phi\rangle^2\ ,\eeq
where the inequality is saturated when there is no other scalar
whose VEV breaks the dark U(1) to give mass to the $Z'$.
Making this assumption, \pref{xdmscatt} implies that 
\beq 
\langle\phi\rangle = 4.6\ \textnormal{GeV}\, 
\left(\frac{\zeta_\downarrow}{3000}\right)^{1/4}
\left(\frac{\eta}{10^{-3}}\right)^{1/4} \left(\frac{\bar Y}{0.33}\right)^{1/2}
\ .\label{phivev}
\eeq
For standard scalar potentials with dimensionless couplings no greater
than O(1), $m_\phi\lesssim\langle\phi\rangle$, so we can infer an
upper limit on the mass of $\phi$.
Moreover, the definition of the mixing parameter $\eta$ gives
\beq y\simeq \frac{\sqrt 2\eta M_\chi}{\langle\phi\rangle} =
3.0\times 10^{-3} \, \left(\frac{M_\chi}{10\ \textnormal{GeV}}\right)
\left(\frac{\zeta_\downarrow}{3000}\right)^{-1/4}
\left(\frac{\eta}{10^{-3}}\right)^{3/4}\left(\frac{\bar Y}{0.33}\right)^{-1/2}
\ .\label{phiyukawa}\eeq

Table \ref{parameters} summarizes the values of $\langle\phi\rangle$,
$y$, $m_{Z'}$, and $g$ in terms of the other model parameters.  We
take as fiducial values  $M_\chi=10$ GeV,
$\zeta_\downarrow=3000$,  $\eta=10^{-3}$, and $\bar Y=1/3$. 
These are further supplemented by
\pref{xdmscatt} relating $m_{Z'}$ and $g$ and \pref{lambda3} for the
cubic coupling of the neutral scalar $\Phi$.

\section{Other constraints}\label{constraints}
Besides the relic density and rate of galactic positron production,
we need to insure that the metastable state $\chi_3$ is sufficiently
long-lived and that it is not depleted by
$\chi_3\chi_3\to\chi_2\chi_2$ downscattering in the early universe.

\subsection{Lifetimes of excited states}\label{lifetimes}

A necessary condition for our exothermic XDM model is that the
excited state $\chi_3$ must be long-lived compared to the age of
the universe.  In fact, the rate of 
$\chi_3\to\chi_1 e^+ e^-$ decays must be significantly less than 
that of $\chi_3\chi_3\to \chi_2\chi_2$ downscatterings to insure that
the 511 keV signal has the right morphology, proportional to
the DM density profile squared, $\rho(r)^2$, rather than $\rho(r)$.

The $\chi_2$ state on the other hand 
must decay relatively quickly via $\chi_2\to\chi_1 e^+ e^-$, mediated
by $Z'$ exchange.  Let us consider it first.  
From eq.\ (50) of \cite{CFC}, the rate is given by
\beq
\Gamma_2 = 2\alpha_g\, \alpha\epsilon^2\eta^2m_e^2\, (\delta M_\chi-2m_e)^3 
m_{Z'}^{-4}\simeq \frac{\alpha g^2\epsilon^2\eta^2 m_e^5}{2\pi\, m_{Z'}^4}\ ,
\label{decayratebyZ}\eeq
where $\alpha$ is the fine structure constant, $\alpha_g = g^2/4\pi$, 
and $\epsilon$ is the kinetic mixing parameter between $Z'$ and the photon.   
We have assumed that $\delta M_\chi -2 m_e\sim m_e$ in the
absence of fine tuning.
$\tau_2 = 1/\Gamma_2$ should be small compared to the time it takes a 
$\chi_2$ DM particle to travel $\sim 1$ kpc at the typical speed of 
$10^{-3} c$.  Otherwise the signal will become spatially more spread-out
than observed due to the propagation of $\chi_2$.
This gives the constraint
\beq
\frac{\sqrt{\eta g\epsilon e}}{m_{Z'}} \gg 
\frac{1}{82\textnormal{ TeV}}\ ,
\eeq
which is easily satisfied.  For example using $\eta=10^{-3}$, 
$M_\chi = 10$ GeV, $m_{Z'} = 600$ MeV, $g = 0.09$ (see eq.\
(\ref{relicvals})), we find $\epsilon\gg 1.6\times 10^{-6}$.

The decay $\chi_3\to \chi_1 e^+e^-$ is mediated by the $\phi$
and $\Phi$ 
mixing with the SM Higgs.  The matrix element is the same as in the 
vector-mediated case except that both $\phi$ and $h$ contribute, leading to
a decay rate
\beq
\Gamma_3 = \frac{\alpha_{y_e}\, 
m_e^2\, (\delta M_\chi-2m_e)^3}{2\pi} \left[\frac{yc_\theta s_\theta}{m_\phi^2}
-\frac{y_\Phi \eta c_\Theta s_\Theta}{m_\Phi^2} +\frac{y_\Phi \eta c_\Theta s_\Theta
-yc_\theta s_\theta}{m_h^2}\right]^2\ ,
\label{decayratebyphi}	
\eeq
where $\alpha_{y_e} = m_e^2/4\pi v^2$,
$v =246$ GeV is the Higgs VEV, and $\theta,\Theta$ are  the $\phi$-Higgs
and $\Phi$-Higgs mixing angles respectively. 
In the absence of finely-tuned destructive interference between the 
Yukawa couplings and mixing angles, we find
\beq
\Gamma_3\simeq \textnormal{max}\left(
\frac{y^2\, \theta^2\, m_e^7}{8\pi^2\, M_\phi^4\, v^2}\, ,
\frac{y_\Phi^2\eta^2\, \Theta^2\, m_e^7}{8\pi^2\, M_\Phi^4\, v^2}\right)\ .
\label{decayratebyphi2}\eeq
Here we have
assumed that $\theta,\Theta \ll 1$ and denoted 
$M_{\phi,\Phi}^4 = (m_h^{-2} - m_{\phi,\Phi}^{-2})^{-2}$. (Notice that $M_{\phi,\Phi}$ is
approximately given by min($m_{\phi,\Phi}$, $m_h$) if the dark scalar masses
 are not close to that of the SM Higgs.)   Small values of $\theta,\Theta$ are
clearly preferred for suppressing $\Gamma_3$.

Comparing to the observed rate of positron
emission in the galactic center, eq.\ (38) of \cite{CFC}, we can
constrain the couplings in terms of the other model parameters.
Demanding that $\Gamma_3$ be sufficiently small, so that downscatterings
dominate over decays to produce the observed positrons, we find
(in the case where $\phi$ exchange dominates the decay of $\chi_3$)
\beq
	y^2<  \left(\frac{0.33}{\bar Y}\right)
		\left(\frac{25}{\zeta}\right)
	\left(\frac{\delta M_\chi}{m_e}-2\right)^{-3}
	\left(\frac{10^{-2}}{\theta}\right)^2 \,
	\frac{ M_\phi^4\, M_\chi}{(600.3\textnormal{ GeV})^5}
\label{yconst}
\eeq
where $\zeta$ is defined as
$\zeta = \textnormal{kpc}^{-3} \int_0^{r_c} dr\, r^2 \rho(r)/\rho_\odot$
as in eq.\ (38) of \cite{CFC}
(recall that $\bar Y = Y_3/Y_{\rm tot}$ is the abundance of 
$\chi_3$ relative to
the total $\chi$ number).   An analogous bound applies when $\Phi$
dominates in $\chi_3$ decay.  We thus infer that 
\beq\label{yconst2}
\textnormal{max}\left(\frac{\sqrt{y\, \theta}}{M_\phi}\, ,
\frac{\sqrt{y_\Phi \eta \Theta}}{M_\Phi}\right) < \frac{1}{19\textnormal{ TeV}}
\left(\frac{M_\chi}{10\textnormal{ GeV}}\right)^{1/4}
\left(\frac{\zeta}{25}\right)^{-1/4}\left(\frac{\bar Y}{0.33}\right)^{-1/4}
\left(\frac{\delta M_\chi}{m_e}-2\right)^{-3/4}\ .
\eeq
For example, with $M_\phi\simeq m_\phi\simeq 4.6$ GeV and 
$y=3\times 10^{-3}$ as in (\ref{phivev},\ref{phiyukawa}), the
$\phi$-$H$ mixing angle must be very small, 
$\theta \lesssim 2\times 10^{-5}$.  The bound on $y_\Phi \Theta$ is 
less stringent both due to the presence of $\eta$ and the lack of
an upper limit on $m_\Phi$.  With $m_\Phi\sim 5$ GeV, $y_\Phi\sim 1/10$, and
$\eta\sim 10^{-3}$, we find $\Theta\lesssim 6\times 10^{-4}$.
We note that if $m_\phi,m_\Phi\ll m_h$,
then a percent-level tuning of $y\theta m_\phi^{-2}-y_\Phi\eta \Theta m_\Phi^{-2}$
can loosen these constraints significantly.

\comment{This is acceptable in the context of our model
since with $\mu\sim 10$ MeV, the VEV of $\phi$ has a reasonable value,
 $\langle\phi\rangle = \mu/y
\simeq 40$ GeV. }

\subsection{Washout of metastable states}\label{washout}

In order for the relic density of the metastable state $\chi_3$  to be
relatively undepleted, we require that downscattering reactions
capable of changing $\chi_3$ number freeze
out at temperatures above the respective mass splitting 
$\delta M_{23}\simeq$ 10 keV or $\delta
M_{13}\simeq$ MeV.     To be precise, these refer to the 
\textit{kinetic} temperature of the $\chi$ particles, which differs from
the temperature of photons if the DM particles are no longer in 
kinetic equilibrium with the SM thermal bath. 

We first consider kinetic equilibrium with the SM.  The $\chi_{2,3}$ 
population can be held in equilibrium with the SM by
the single scattering process $\chi_3 e\leftrightarrow \chi_2 e$,
which was considered for similar GeV-scale XDM in \cite{CFC},
with decoupling temperatures given in figure 11 of that 
reference.\footnote{The $Z'$ fine structure constant has approximately
the same functional dependence on $M_\chi$ as $\alpha_g$ in \cite{CFC}.}
For the preferred values of $m_{Z'},g,\epsilon$ given for direct detection 
in section \ref{direct} below, decoupling from the SM occurs at 
$T_d=36$ MeV; however, as indicated in section \ref{direct}, scenarios with
larger values of $\epsilon$ and therefore lower $T_d$ are possible.
Since the kinetic temperature is $T_k=T^2/T_d$ for SM temperature $T$,
we  have $T_k=\delta M_{13}$ at $T\sim 5\,\delta M_{13}$ (assuming
$\delta M_{13}-2m_e\sim m_e$) and $T_k=\delta M_{23}$ at $T\sim 60\,\delta M_{23}$
(for $\delta M_{23}\sim 10$ keV).  The following 
constraints become less stringent if kinetic decoupling from 
the SM occurs later than 10 MeV temperatures.

Since we are concerned with the relic abundance of $\chi_3$, we need to
ask which processes will change $\chi_3$ number most efficiently.
Besides scattering from electrons, $\chi_3$ will be held in kinetic 
equilibrium by rapid $\chi_3\chi_3\leftrightarrow\chi_2\chi_2$ scattering
mediated by $Z'$.  The dominant 
scalar-mediated processes are $\chi_3\chi_3\leftrightarrow \chi_1\chi_1$
and $\chi_3\chi_1\leftrightarrow \chi_1\chi_1$.  
There is no threshold
velocity for downscattering, so the value of
$\langle\sigma_{\downarrow} v_{\mathrm{rel}}\rangle$ at $v=0$ is the relevant one.
For the three downscattering processes, these cross sections are
\beqa
\langle\sigma_{33\to 11} v_{\mathrm{rel}}\rangle &=& 
\frac{ M_\chi^2}{4\pi}\left(\frac{y^2}{m_{\phi}^2}-\frac{y_\Phi^2\eta^2}{m_\Phi^2}
\right)^2 \sqrt{\frac{2\delta M_{13}}{M_\chi}}\ ,\nonumber\\
\langle\sigma_{31\to 11} v_{\mathrm{rel}}\rangle &=& 
\frac{ M_\chi^2}{4\pi}\left(\frac{y^2\eta}{m_{\phi}^2}-\frac{y_\Phi^2\eta}{m_\Phi^2}
\right)^2 \sqrt{\frac{\delta M_{13}}{M_\chi}}\ ,\nonumber\\
\langle\sigma_{33\to 22} v_{\mathrm{rel}}\rangle &=& 
\frac{ M_\chi^2}{4\pi}\left(\frac{g}{m_{Z'}}\right)^4
\sqrt{\frac{2\delta M_{23}}{M_\chi}}\ .
\label{sigdown}\eeqa
As throughout, we 
have assumed that the mediator masses satisfy $m^2\gg M_\chi\delta M$ 
for the mass splitting of each process above.  We have ignored
the $s$-channel in scalar-mediated processes because it is suppressed
by powers of velocity.

Since we are finding that downscattering freezes out before
depleting the excited states, the DM states have number density
\beq
n_3\simeq T^3\, \bar Y\left(\frac{\xi}{M_\chi}\right)  
\left(\frac{g_{*S}(T)}{g_{*S}(T_0)}\right)\ , \ \  
n_1\simeq T^3 \left(\frac{Y_1}{Y_{tot}}\right)\left(\frac{\xi}{M_\chi}\right)
 \left(\frac{g_{*S}(T)}{g_{*S}(T_0)}\right)\ ,\eeq
where
$\xi=7\times 10^{-10}$ GeV.  We estimate that freezeout occurs at 
$\Gamma_\downarrow \simeq 3H$, with $H\simeq 1.66\sqrt{g_*(T)}$ $T^2/M_P$,
which should happen before $e^+e^-$ annihilation.
For $\chi_3\chi_3\to \chi_1\chi_1$ scattering,
\beq T_f = 153 \left(\frac{y^2}{m_\phi^2}-
  \frac{y_\Phi^2\eta^2}{m_\Phi^2}\right)^{-2}
  \frac{1}{\xi M_P \sqrt{M_\chi\delta M_{13}}} 
	\left(\frac{0.33}{{\bar Y}^{\phantom{1^1}\!\!\!\!\!}}\right)
\ .
\eeq
Requiring $T_f\gtrsim 5\,\delta M_{13}$, this generically 
leads to the constraint
\beq \textnormal{max}\left(\frac{y}{m_\phi},\frac{y_\Phi\eta}{m_\Phi}\right)
< \frac{1}{15\textnormal{\ GeV}}
\left( \frac{10\textnormal{ GeV}}{M_\chi}
\right)^{1/8} \left(\frac{0.33}{\bar Y}\right)^{1/4}
\left(\frac{3m_e}{\delta M_{13}}\right)^{3/8}
\ ,\eeq
which is easily compatible with the fiducial values $y\sim 3\times 10^{-3}$,
$m_\phi\sim 4.6$ GeV.  It is also compatible with any perturbative value of 
$y_\Phi$ for $m_\Phi\gtrsim$ GeV and $\eta\sim 10^{-3}$.
Similarly, for $\chi_3\chi_1\to \chi_1\chi_1$ downscattering, we find
\beq 
\textnormal{max}\left(\frac{y\sqrt\eta}{m_\phi},\frac{y_\Phi\sqrt\eta}{m_\Phi}
\right) < \frac{1}{14\ \textnormal{GeV}} 
\left( \frac{10\textnormal{ GeV}}{M_\chi}
\right)^{1/8} \left(\frac{0.33}{Y_1/Y_{tot}}\right)^{1/4}
\left(\frac{3m_e}{\delta M_{13}}\right)^{3/8}\ ,
\eeq
which is likewise a weak constraint.  Finally, $\chi_3\chi_3\to\chi_2\chi_2$
freezeout (which occurs after $e^+e^-$ annihilation) yields 
\beq 
\frac{g}{m_{Z'}} <\frac{1}{3.8\ \textnormal{GeV}}\left(
\frac{10\ \textnormal{GeV}}{M_\chi}\right)^{1/8}
\left(\frac{0.33}{\bar Y}\right)^{1/4}
\left(\frac{10\ \textnormal{keV}}{\delta M_{23}}\right)^{3/8}
\ .\eeq
This is compatible with (\ref{xdmscatt}) for our fiducial
parameter values, in particular as long as 
$\zeta_\downarrow > 350$, which is a very mild lower limit on the 
cuspiness of the DM density profile (fig.\ \ref{zetap}) compared to that
expected from $N$-body simulations.

\subsubsection*{Kinetic before chemical freezeout?}

We have so far assumed that $\chi_1$ is completely decoupled from
$\chi_{2,3}$ by the chemical freezeout, which necessitated
a separate 
annihilation channel for $\chi_1$.  If $\chi_{2,3}$ can efficiently
downscatter into $\chi_1$ during chemical freezeout, then the relic 
abundance for all three DM species is determined simply by the 
annihilation and coannihilation of $\chi_2$ and $\chi_3$ with subsequent
scattering into $\chi_1$.  In this case, the neutral $\Phi$ scalar 
is not necessary and could be eliminated from the model.

Under what conditions could this simplification be achieved?  
In the absence of the $\Phi$, the scattering processes 
$\chi_3\chi_3\leftrightarrow\chi_1\chi_1$ and 
$\chi_2\chi_3\leftrightarrow\chi_2\chi_1$ are parametrically 
the largest among
scatterings mediated by $\phi$ and $Z'$ respectively.
For them to be in equilibrium during chemical freezeout
at $T\simeq M_\chi/18$, we require
\beqa 
   \frac{y}{m_\phi} &>& \frac{1}{54\ \textnormal{GeV}} 
   \left( \frac{10\textnormal{ GeV}}{M_\chi}
   \right)^{3/8} \left(\frac{0.33}{\bar Y}\right)^{1/4}
   \left(\frac{3m_e}{\delta M_{13}}\right)^{1/8}\ \textnormal{or}\nonumber\\
   \frac{g\sqrt\eta}{m_{Z'}} &>& \frac{1}{59\ \textnormal{GeV}} 
   \left( \frac{10\textnormal{ GeV}}{M_\chi}
   \right)^{3/8} \left(\frac{0.33}{\bar Y}\right)^{1/4}
   \left(\frac{3m_e}{\delta M_{13}}\right)^{1/8}\ .
\eeqa
The first is clearly ruled out for the parameter ranges of interest.
In light of equation \pref{xdmscatt}, the second is also difficult to 
achieve but may occur for large $\eta$ corresponding to 
$\delta M_{23}\simeq$ MeV, at the boundary of the allowed parameter space.
Apart from this marginal possibility, kinetic freezeout of $\chi_1$
generically occurs before chemical freezeout.

\subsection{Astrophysical Constraints}\label{astroconstraints}

Light (GeV to 10 GeV mass) dark matter is subject to a number of 
constraints from astrophysics, which we review briefly here as applied
to our model.

The cosmic microwave background (CMB) begins to constrain the possible
annihilation channels of light DM.  These can inject energy into the SM plasma at 
redshifts from approximately 100 to 1000, slightly modifying 
recombination; an inexhaustive list of recent references includes
\cite{Galli:2011rz,Hutsi:2011vx,Finkbeiner:2011dx}.  Current limits
disfavor DM annihilations primarily to electrons for DM masses
$M_\chi \lesssim 10$ GeV with the standard thermal cross section, but 
annihilation 
primarily to muons (or to any other particle whose decay products contain
significant numbers of neutrinos) 
is  allowed \cite{Galli:2011rz,Hutsi:2011vx}.
For $Z'$ masses considered here, our models have a branching ratio of
about 44\% into $4\mu$ and another 44\%
into $2\mu,2e$ after decay of the $Z'$ in the $\chi\chi\to Z'Z'$ branch
As a result, they are subject to the more relaxed limits.  Furthermore, because 
$Z'$ is relativistic at thermal freezeout, the annihilation cross
section is slightly reduced compared to the canonical value, relaxing the
constraints somewhat.  In addition, ref.\ \cite{Steigman:2012nb} has recently 
argued that the canonical value of the thermal abundance cross section 
should itself be decreased in the vicinity of $M_\chi=10$ GeV, again loosening
CMB bounds.  As a result, our models evade these constraints at present.
The effects on the CMB of $e^+$ production via downscattering are 
currently under investigation \cite{freyreid}.

Diffuse gamma ray emission in our galaxy can similarly be used to place 
limits on annihilation of DM to SM fermions via final state radiation
and inverse Compton scattering.
(see, for example \cite{Ackermann:2012rg}, for current results).  
However, these limits are both
weaker than those extracted from the CMB and more subject to astrophysical 
uncertainty.

Dark matter can also affect stellar evolution if it has a large enough
capture cross section to accumulate significantly within stars (see
\cite{Scott:2008ns} and references therein).  Only stars orbiting
close to the center of our galaxy would be able to capture enough
WIMPs to undergo significant effects.  With sufficient accumulation, the
luminosity from DM annihilations can suppress nuclear burning and
increase the star's main-sequence lifetime, with an effect going
inversely to the mass of the star. In \cite{Scott:2008ns} it was shown
that for spin-dependent DM-nucleon scattering cross section
$\sigma_{SD}=10^{-38}$ cm$^2$ and mass $M_\chi =100$ GeV, and assuming
a cuspy DM density profile, these effects could be observable in
binary systems in close elliptical orbits around the galactic center.  The figure
of merit for an observable effect is the ratio of the capture cross
section to $M_\chi$.  For spin-independent scattering, capture on
helium is $\sim 10$ times more effective than spin-dependent
capture on hydrogen, but only $\sim 10/4$ for a model like ours where DM
interacts with protons but not neutrons.  Thus to be on the
borderline for an observable effect, we would need $\sigma/M_\chi
\gtrsim 4\times 10^{-41}$ cm$^2$. In the examples we will discuss in
section \ref{direct}, this criterion is satisfied, and so one could
hope to observe binary systems with unusual properties near the
galactic center. But at the present, given the great uncertainties in
the DM density at the galactic center (in a much smaller region than
that over which the positron annihilation for the 511 keV signal is
occurring), such considerations cannot yet place a constraint on the
model, especially since \cite{Scott:2008ns} assumed an NFW profile to
obtain their optimistic results, which is more cuspy than the Einasto
profile we have assumed.

\section{Direct detection}\label{direct}

There has been much interest in hints of light dark matter from the
DAMA \cite{Bernabei:2008yi}, CoGeNT \cite{Aalseth:2010vx} and  CRESST
\cite{Angloher:2011uu} experiments, supported by evidence of annual
modulations in the first two \cite{Aalseth:2011wp}.   It is difficult
to engineer models with the right properties to fit all sets of
observations, but less for only one of them, in
particular that of CoGeNT (see for example \cite{Farina:2011pw}).    
The best fit for CoGeNT alone, according to the collaboration's preliminary estimates
that take into account contamination due to surface events
\cite{collar} is $M_\chi\simeq 10$ GeV and $\sigma_n = 2\times
10^{-41}$ cm$^2$.   This region is excluded by Xenon100
\cite{Aprile:2011hi}, but the edge of the CoGeNT-allowed region at
$M_\chi = 7$ GeV,  $\sigma_n = 5\times 10^{-41}$ cm$^2$ is marginally
compatible; we will focus on these values.  In a model like ours where
DM interacts only with protons, one must rescale the cross section by
$(A/Z)^2$ which for Ge is a factor of 5.2, so that the desired cross
section on protons is  $\sigma_p = 2.6\times 10^{-40}$ cm$^2$. 

The cross section for $\chi_{3}$ scattering on protons, mediated by
$Z'$ exchange, is given
by
\beq
\label{sigmap}
\sigma_p = \frac{(g\epsilon e \mu_N)^2}{\pi m_{Z'}^4}
\eeq
where $\mu_N$ is the DM-nucleon reduced mass.  
Using $M_\chi = 7$ GeV in (\ref{relicvals}), corresponding to the lowest (and least
fine-tuned) relic density contour of fig.\ \ref{relicS}, we find that
$m_{Z'} = 264$ MeV and $g = 0.047$.  Then (\ref{sigmap}) gives the
desired value of $\sigma_p$ to match CoGeNT if
\beq
	\epsilon \simeq 10^{-6}
\eeq
Unfortunately it is difficult to find further complementary evidence
for such a weak level of kinetic mixing.  It is an order of magnitude below
the sensitivity of the Heavy Photon Search (HPS) beam dump
experiment proposed at Jefferson Laboratory \cite{HPS}.

However, a DM particle with slightly lower mass $M_\chi\sim 4$ GeV
and larger cross section $\sigma\sim 5\times 10^{-39}$ cm$^2$ 
can have a larger value of $\epsilon$.  These numbers correspond to
the best fit for CoGeNT found by ref.\ \cite{Farina:2011pw} for the
case of 10\% channeling (see fig.\ 13 of that paper), taking into
account our $(A/Z)^2$ correction factor.  Although such
a large channeling fraction is not a favored scenario 
\cite{Bozorgnia:2010ax}, we choose it to illustrate that such values
are within reason.  In this case, (\ref{relicvals}) gives 
$m_{Z'} = 200$ MeV, $g = 0.036$.  Then $\epsilon = 1.5\times 10^{-3}$,
which is in fact already marginally excluded by recent limits from 
the MAMI (Mainz Microtron) \cite{Merkel:2011ze} and APEX 
\cite{Abrahamyan:2011gv} experiments.  Clearly then, the value of $\epsilon$
needed to saturate DM direct detection limits is rather sensitive to
$M_\chi$ and there is room for models in which not only the DM itself
but also the new mediator particle can be directly observed.

We note that any relic population of $\chi_1$ particles, even if
comparable in number to $\chi_{2,3}$, will
be undetectable by current direct detection experiments.  First, only 
the elastic scattering $\chi_1 N\to \chi_1N$ mediated by $\Phi$ is 
kinematically allowed, and this has cross section (on nucleons)
\beq\label{sigmaN}
\sigma_N=\frac{(y_\Phi\Theta y_N\mu_N)^2}{\pi m_\Phi^4}\ ,\eeq
where $y_N=0.3 m_N/v\approx 1.14\times 10^{-3}$ is the nucleon-Higgs
Yukawa coupling ($v=246$ GeV is the Higgs VEV).  For $M_\chi=7$ GeV
and fiducial
values for the other parameters as given in table \ref{parameters}, this is
$\sigma_N=6.3\times 10^{-46}$ cm$^2$, several orders of magnitude below 
current experimental
limits.

\section{Conclusions}

In this work we have explored the consequences of assuming that
the excess galactic 511 keV signal is results from heavy (compared to the
MeV scale) dark matter interacting via an abelian gauge interaction.
While it will be difficult to demonstrate that this signal is due to 
dark matter rather than some more mundane astrophysical mechanism,
the possibility of having independent discovery of elements of such a model
from direct detection of the dark matter or of the dark gauge boson 
could make such an identification more plausible.   (Higher resolution
measurements of the morphology of the 511 keV signal by a future 
low-energy $\gamma$-ray mission would also help to settle this question.) 
Conversely, the ability to exclude part of the parameter space by independent
experiments is an interesting feature of this type of model.
The model we have presented with the fiducial parameter values listed
in table \ref{parameters} gives an existence proof that abelian DM
can produce enough positrons to explain the 511 keV signal as well
as avoid known constraints and potentially provide independent signals.

This model, having three components of dark matter and two new Higgs bosons,
is surprisingly rich, including some  complications necessitated by the many
constraints which must be satisfied: naturally small mass splittings, a
long-lived excited state,  short-lived intermediate state, and correct relic
densities of the three DM species .  Perhaps the most peculiar feature of the
model is an accidental near-degeneracy of the DM mass parameters $M_\chi$ and
$m$ such that $M_\chi -m \gtrsim 2 m_e$.  It should be emphasized that this 
is not a fine-tuning in the usual sense since the fermion masses are
protected by chiral symmetry and therefore the MeV-scale splitting is not
destabilized by radiative corrections.  A more complete theory would
presumably give an explanation for this mass difference, just as the eventual
successor to the standard model of particle physics will hopefully explain
why $m_e$ is much smaller than the mass of the top quark.  

As a compensation for its lack of elegance, we find that the model can have
interesting  predictions for direct detection in the regions suggested by
CoGeNT and CRESST (or lower masses $\sim 4$ GeV), and possibly for beam-dump
searches for light vector bosons such as the APEX, MAMI and HPS experiments.
The dark Higgs boson $\Phi$ could moreover mix with the standard model Higgs
boson and manifest itself through invisible decays of the latter.

\begin{acknowledgement}
This work was supported in part
by the Natural Sciences and Engineering Research Council of Canada.
\end{acknowledgement}

\bibliographystyle{adp}
\bibliography{abelianxdm}

\end{document}